\documentclass[review,journal]{IEEEtran}    

\usepackage{graphicx}
\usepackage{float}
\usepackage{placeins}
\usepackage{amsmath}
\usepackage{booktabs}
\usepackage{tabu}
\usepackage{CJKutf8}      
\usepackage{xcolor}
\usepackage{subcaption}

\usepackage[hidelinks]{hyperref}

\begin{document}

\title{PatternSight: A Perceptual Grouping Effectiveness Assessment Approach\\for Graphical Patterns in Charts}

\author{%
  Xumeng~Wang*,
  Xiangxuan~Zhang, 
  Zhiqi~Gao, 
  Shuangcheng~Jiao, 
  and~Yuxin~Ma
  
  \thanks{X. Wang, X. Zhang, Z. Gao, and S.Jiao are with DISSec, Nankai University.} 
  \thanks{Y. Ma is with Department of Computer Science and Engineering, Southern University of Science and Technology.}
  \thanks{X. Wang is the corresponding author. E-mail: wangxumeng@nankai.edu.cn.} 
}

\markboth{IEEE Transactions on Visualization and Computer Graphics,~Vol.~XX, No.~XX, Month~Year}%
{Wang \MakeLowercase{\textit{et al.}}: PatternSight}
\maketitle

\begin{abstract}
The boom in visualization generation tools has significantly lowered the threshold for chart authoring. Nevertheless, chart authors with an insufficient understanding of perceptual theories may encounter difficulties in evaluating the effectiveness of chart representations, thereby struggling to identify the appropriate chart design to convey the intended data patterns. To address this issue, we propose a perception simulation model that can assess the perceptual effectiveness of charts by predicting graphical patterns that chart viewers are likely to notice. The perception simulation model integrates perceptual theory into visual feature extraction of chart elements to provide interpretable model outcomes. Human perceptual results proved that the outcome of our model can simulate the perceptual grouping behaviors of most chart viewers and cover diverse perceptual results. We also embed the model into a prototype interface called PatternSight to facilitate chart authors in assessing whether the chart design can satisfy their pattern representation requirements as expected and determining feasible improvements of visual design. According to the results of a user experiment, PatternSight can effectively assist chart authors in optimizing chart design for representing data patterns.
\end{abstract}

\begin{IEEEkeywords}
Human perception simulation, perceptual grouping, graphical pattern identification.
\end{IEEEkeywords}


\section{Introduction}

Due to the enhanced capability for information representation, visualizations are ubiquitous in contexts where individuals seek to illustrate data patterns and derive corresponding insights. Visualizations help achieve this by transforming data records into graphical elements that employ various \textit{visual effects}--the rendering effectiveness of visual channels, such as color, shape, size, etc. A group of coordinated graphical elements can be identified as \textit{graphical patterns} through perceptual grouping~\cite{wagemans2012century} and represent data patterns effectively. Taking advantage of authoring tools (e.g., voyager 2~\cite{wongsuphasawat2017voyager}, NL interpreter~\cite{wang2022towards}, MARVisT~\cite{chen2019marvist}, Data Illustrator~\cite{liu2018data}, Falx~\cite{wang2021falx}, Tableau, and PowerBI), individuals with limited expertise in visualization can create graphical patterns and become authors of visualizations~\cite{rubab2021examining} by taking recommended design schemes or employing chart templates.

The process of authoring visualizations to represent a specific data pattern is deceptively simple. 
With the absence of prior knowledge about the data patterns, visualization viewers may rely on the visual stimuli they receive to identify visual elements as unitary organizations, in other words, graphical patterns. however, the graphical patterns that are perceived in priority may not correspond to the data patterns emphasized by chart authors~\cite{quadri2024you}. Without formal training, authors may lack an understanding of the principles of perceptual grouping, inadvertently producing charts that highlight unintended information and distract the attention of viewers. When confronted with complex data challenges, such as dense distributions or multiple dimensions~\cite{munzner2009nested,bresciani2015pitfalls,qu2017keeping,lo2022misinformed}, even experienced designers may encounter similar pitfalls. Consequently, there is a pressing need for visualization authors to assess potential perceptual outcomes.

A detailed assessment of the perceptual grouping effectiveness of visualizations requires conducting experiments or interviews with human participants, who serve as potential representative viewers~\cite{lam2011empirical, elliott2020design}. However, implementing such an assessment process can be labor-intensive or impractical for authors who utilize visualizations for convenience. The authoring process can benefit from rapid assessment or timely advice, which suggests the need for an automated assessment approach. Nevertheless, simulation of visual perception presents significant challenges as the human visual perceptual system is far more complex than the architecture of artificial intelligence. In addition, annotating visualizations for perception grouping effectiveness requires annotators to summarize the graphical patterns they perceive and the intensity of perception after browsing the material. Such requirements of detailed annotations can impede large-scale artificial intelligence models from collecting adequate training data. 

The theories of visual perception have been explored for decades and employed to guide visualization design~\cite{quadri2021survey, gramazio2014relation}. Nevertheless, there is still a research gap between perceptual grouping theories and practices of pattern representation. To fill this gap, we summarize visual effects that may affect graphical pattern identification based on existing perceptual grouping theories and propose a perception simulation model to refine grouping rules from human annotations. The proposed model can be applied to identify graphical patterns that are likely to be recognized by a majority of viewers, while also suggesting a range of perceptual outcomes. To support model application, we further develop PatternSight, a prototype interface that provides chart authors with explanations of the identified patterns based on visual effects and provides chart design suggestions. The effectiveness of the perception simulation model and the interface is evaluated through a performance test and a user study to demonstrate their usability in enhancing visualization authoring processes.


\section{Related Work}


\subsection{Creation Support from Visualization Authoring Tools}
Visualization authoring tools trade expressivity of visualizations with the cost of usability~\cite{satyanarayan2019critical}. In addition to the feature of visualization generation, authoring tools need to provide authors with design support to ensure cost-effectiveness. To this end, existing studies have made efforts from two aspects: improving the quality of built-in visualization templates and giving feedback on visualization design. 

Existing visualization authoring tools support design in various ways: Matplotlib provides low-level plotting interfaces, emphasizing flexibility but requiring manual configuration of visual encodings; Altair, Vega, and Vega-Lite are based on declarative syntax, simplifying visualization generation through templates and constraints; Draco recommends visualization designs conforming to perceptual principles through logical rules. While these tools excel in template generation or rule recommendation, they lack dynamic assessment of the perceptual effectiveness of charts-i.e., predicting graphical patterns that viewers are likely to perceive, which is the core difference of this study.

Applying templates is the most convenient approach to visualization creation for non-professional authors. In search of high-quality built-in visualization templates, authoring tools need to first collect well-designed visualizations. Bako et al.~\cite{bako2022streamlining} use D3 examples as a visualization source and gather templates for common chart types, such as bar charts and force-directed graphs. However, collecting sufficient visualization templates to cover the entire design space and chart types exhaustively is challenging for authoring tools. To break the limitation caused by the number of available templates, authoring tools also need to support creating visualizations beyond built-in templates~\cite{way2018unconventional, wootton2024charting}. 

When specifying visualization details, authors require feedback from authoring tools, like recommendations on feasible choices or reminders of underlying errors. To summarize feasible choices, authoring tools can build a corpus. For example, given a corpus on the effectiveness of annotation in visualizations, authoring tools~\cite{sultanum2021leveraging, stoiber2023authoring} can recommend annotation styles for specified visualization objects. MultiVision~\cite{wu2021multivision} takes metrics such as chart diversity into consideration and recommends charts for dashboards from a visualization corpus. To identify design errors and make reminders, authoring tools need to find ``standard answers'' for visualization design. Dupo~\cite{kim2023dupo} is designed to produce edit suggestions such as ``adding ticks for axes'' according to the comparison results between the current visualization and similar works. Authors can interactively adopt the suggestions from Dupo through clicking interactions. Suggestions extracted from high-quality visualization works are worth learning from but lack explanation. It is difficult for authors to follow suggestions blindly without making the same mistakes again. Besides chart improvement suggestions, PatternSight delivers underlying perceptual results and explains how the visual effects of graphical elements affect pattern perception. Feedback collected from the user study confirms that such explanations help chart authors understand perceptual mechanisms and gain confidence when optimizing charts.

\subsection{Information Extraction from Visualization}
Extracting information from visualization is necessary for models to understand or assess visualization. The information used as model input mainly falls into visual information, data information, and structural information. 

Visual information~\cite{gupta1997visual} refers to visible objects or concepts, which is a research focus in computer vision (CV). Instead of what can be perceived by humans, visual information concerns what is included in images. CV models are trained to understand visual information in images. The performance of CV models can be verified by Q\&A tests~\cite{shah2019kvqa}. Models~\cite{qiao2018exploring} that learn from visual perception behaviors of humans have a chance to achieve better performance. Considering visualizations as image data, visual information extracted from visualizations can also be used to answer questions~\cite{kafle2020answering}. 

Unlike images, visualizations carry data information that aims to represent data patterns. Visualizations in the format of Scalable Vector Graphics (SVG) also embed structural information of graphical elements. Models can take advantage of extra information when learning from visualizations. For instance, ChartQA~\cite{masry2022chartqa} integrates data tables with visual information to answer questions on visualizations. Li et al.~\cite{li2022structure} generate embedding vectors for SVG-based visualizations based on both structural information and visual information included in the bitmaps that render the corresponding visualizations. The generated vectors record the perceptual effectiveness of visualizations and therefore support quantifying perceptual similarity between visualizations. 

To assess the perceptual effectiveness of graphical patterns in visualizations, our work focuses on the visual effects of graphical elements and how multiple elements form patterns and are noticed by viewers.

\subsection{Perceptual Grouping Theories Applied in Visualization}
A common goal of visualization authoring is to assist viewers in understanding data patterns by perceiving groups of visual elements as graphical patterns~\cite{yi2008understanding}. Gestalt principles stand out from perceptual grouping theories as fundamental references in visualization design~\cite{rosli2015gestalt} and have been validated through user studies on various visualizations~\cite{ventocilla2020comparative}. Fox~\cite{fox2023theories} further synthesized theories and models in graph comprehension, highlighting that effective visualization design must align with how humans cognitively process and interpret graphical elements.


Visualizations for non-specific groups can refer to commonalities summarized in Gestalt principles. Visualization design applying any visual element or visual encoding can be guided by Gestalt principles, especially principles of proximity, similarity, and closure~\cite{rosli2015gestalt}. For example, position encoding is employed by dimensionality reduction algorithms to project individuals. Based on the principle of proximity, Ventocilla et al.~\cite{ventocilla2020comparative} discussed how static results from different projection algorithms support cluster analysis. An incremental projection approach~\cite{fujiwara2019incremental} is proposed to maintain the proximity of individual positions in dynamic projection. The results of the incremental projection approach allow viewers to follow the evolution of individuals. Color encoding considers color discriminability as an assessment metric according to the principle of similarity~\cite{gramazio2016colorgorical, silva2011using}.

Gestalt-based guidelines are summarized separately by data analysis tasks or chart types~\cite{quadri2021survey, zeng2023review}. Zeng et al.~\cite{zeng2023too} introduced an idea of how to recommend a variety of visualizations based on perception theories by training machine intelligence. Training data is indispensable for the practice of the idea, but suitable training data is not yet available. The approach proposed in this paper supports generating universal guidelines that can facilitate visualization optimization across diverse chart types. Besides, PatternSight also supports task-driven visualization optimization by enabling interactive selection of the element groups to be highlighted.

\section{Perceptual Grouping for Graphical Elements}

In this section, we introduce perceptual grouping theories as the background of our approach, synthesize the visual effects mentioned by these theories, and generalize how visual effects are considered in these theories.

\subsection{Perceptual Grouping Theories}
Although a number of theories~\cite{wagemans2012century} have been proposed for describing perceptual grouping, those theories correlate with pattern perception from visualization in different ways. For example, among Gestalt principles, an existing study~\cite{rosli2015gestalt} only highlights principles of proximity, similarity, and closure for visualization design. Different from the first two principles describing how humans group visual elements, the closure principle works when humans recognize a graphical pattern. To explore perceptual grouping from visualization, we summarize perception theories highly relevant to our focus as follows.



\textbf{Proximity}~\cite{izakson2020proximity}: The principle of proximity illustrates how the distance between visual elements affects the result of perceptual grouping. Visual elements that are located closely/next to each other are inclined to be considered a group. Proximity is a dominant principle among Gestalt principles---it could override other principles. A typical application of the proximity principle is scatter plots that show scatters with close locations as a cluster. On the contrary, a scatter far from others is considered an independent individual, or say an outlier.

\textbf{Similarity}~\cite{pinna2022similarity}: Visual elements that share similar values in one or more visual effects are prone to be considered a group. For instance, color is a common method to encode categorical information. Humans naturally regard elements with the same color as belonging to the same category. Nevertheless, the bond of similarity could be impaired or broken by diverse values in other visual effects.

\textbf{Connectedness}~\cite{palmer1994rethinking}: Connectedness represents a special case of proximity. When visual elements share boundaries (i.e., their distance equals zero), they are more readily perceived as a group. For example, nodes in node-link diagrams can be grouped by connected links.

\textbf{Common region}~\cite{palmer1992common}: Elements that share the same region are prone to being considered as a group. In composite charts, like scatter matrices, each chart is assigned an independent cell for pattern representation.

\subsection{Visual Effects and Grouping Considerations}
\label{sec:ve}
The above perceptual grouping theories mainly emphasize \textbf{position} effects and \textbf{appearance} effects.

\textbf{Position}: The principles of proximity, continuity, and common region respectively address the distance between visual elements, their connectivity, and whether they are contained within the same enclosing element. The above conclusions are derived from the position of each element and recorded as pairwise relationships. When evaluating inter-element relationships, closed shapes typically consider boundary boxes, whereas lines or curves focus on the positions of endpoints. Additionally, the similarity principle incorporates assessments of boundary or center position resemblance.


\textbf{Appearance}: The principle of similarity also considers the resemblance of appearance effects, consisting of \textit{type}, \textit{size}, \textit{fill}, and \textit{stroke}. \textit{Types} of visual elements can be divided into shapes (e.g., circles, rectangles) and others (e.g., characters, curves) depending on whether they are closed or not. All visual elements can be displayed with \textit{strokes} and \textit{size}, while \textit{fill} is dedicated to effects of closed elements. The diversity of \textit{stroke} and \textit{fill} can be reflected mainly by color settings (e.g., \textit{hue}, \textit{saturation}, and \textit{lightness}). Closed shapes can vary in \textit{size} across three dimensions: width, \textit{height}, and \textit{area}. Besides, visual elements with outlines may differ in \textit{stroke width}.

\section{Our Approach}

Our approach aims to help chart authors assess and improve the representation effects of graphical patterns in charts. In this section, we describe the details of the design goals and introduce how our approach achieves these goals.

\begin{figure*}[htp]
    \centering
    \includegraphics[width=1.7\columnwidth]{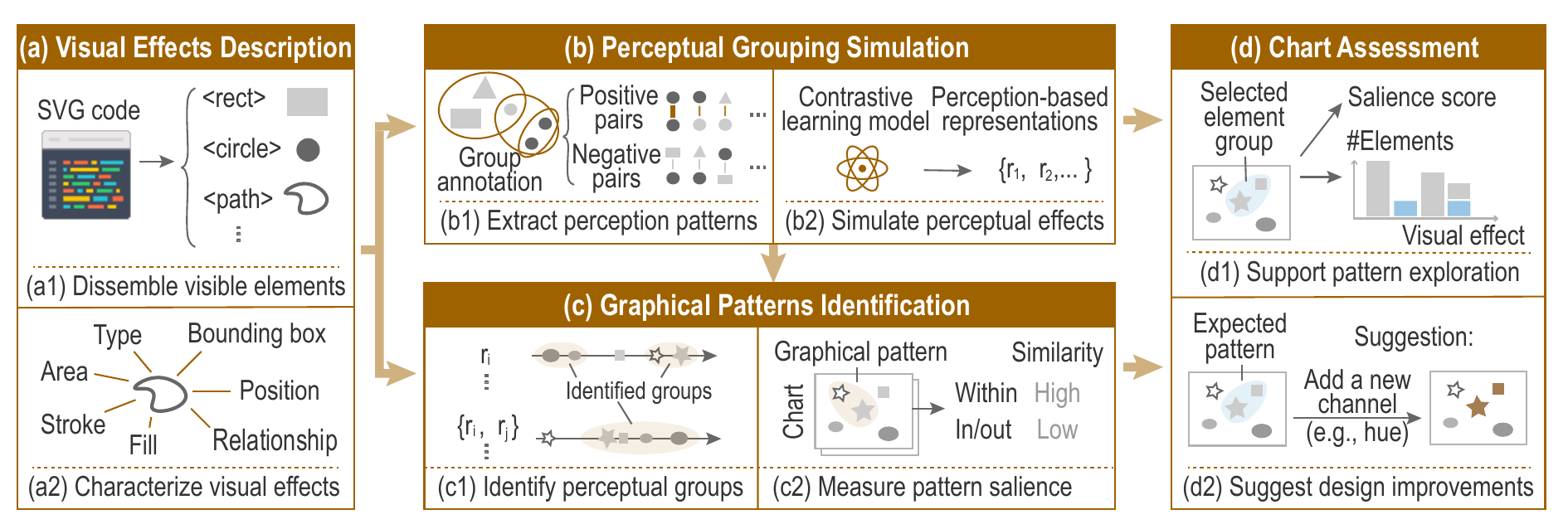}
    \caption{Our approach consisting of 4 parts: (a) describing visual effects of visual elements, (b) formulating a perceptual grouping simulation model based on human annotation results, (c) identifying graphical patterns from charts, and (d) assessing the representation effectiveness of patterns in a chart.}
    \label{fig:pip}
\end{figure*}
\subsection{Design Goals}
We determine the following design goals according to underlying use cases and related studies.

\textbf{G1: Identify graphical patterns based on visual perception of humans.} Although computer vision-based identification approaches can detect graphical patterns from chart images, their ability to reflect how these patterns are perceived by human viewers remains unclear and lacks interpretability. Seeking an accurate and detailed understanding of the representation effects of graphical patterns, the identification process needs to simulate human visual perception behaviors.

\textit{G1.1: Simulate common perceptual grouping behaviors.} Human perception varies across individuals, leading to divergent interpretations of visual stimuli~\cite{reeder2017individual, xu2024insights}. Nevertheless, considering all underlying results could be trivial and time-consuming for chart authors. An efficient solution is to prioritize the perceptual behaviors of the majority. Thus, the perception simulation model should emphasize common perceptual grouping.

\textit{G1.2: Include diverse perceptual results.} Humans can notice multiple graphical patterns from a chart~\cite{bearfield2024same}. While identifying the most salient pattern provides useful insights, relying solely on a single pattern risks oversimplification. Therefore, preserving diversity in graphical pattern identification remains essential for comprehensive perceptual analysis.

\textbf{G2: Explain the perceptual grouping mechanism.} In the design process, feedback on potential effectiveness based on perception mechanisms can better motivate design optimization than absolute judgments, such as ``the current design is invalid''~\cite{rosenholtz2011predictions}. We need a transparent perception simulation mechanism to improve the credibility of its outputs~\cite{shin2021effects}.

\textit{G2.1: Correlate visual effects with graphical patterns.} The perception mechanism is complicated for chart authors lacking specialized training in visual perception. In contrast, explanations grounded in observable visual effects offer more accessible guidance. To generate such explanations, we need to identify and describe the correlations between the visual effects of graphical elements and graphical patterns composed of those elements.

\textit{G2.2: Compare visual effects of elements.}  A feasible approach to understanding the mechanism of a complicated rule is to check the situations when the rule takes effect and when it fails. Grouping rules for graphical patterns can also provide an instantiated explanation for perception mechanisms. To comprehend grouping rules for graphical elements, chart authors can explore which elements could or could not be perceived as a pattern by comparing elements inside and outside each group of elements.

\textbf{G3: Suggest improvements for visual representation.} According to the assessment of graphical patterns, chart authors can assess the representation effectiveness of applied visual encodings and identify potential enhancements. Our tool can facilitate chart authors by suggesting improvement schemes.

\textit{G3.1: Analyze the current representation effects.} Chart authors need to identify the discrepancies between the existing visual effects and the desired effectiveness of pattern representation. For this goal, our tool should provide chart authors with an overview of the current visual effects.

\textit{G3.2: Recommend visual design for representing user-specified patterns.}
There is no best visual design for data visualization, but certain visual design may be superior in representing certain data patterns~\cite{zeng2023review,reimers2024bars}. To generate appropriate recommendations, we should first figure out what data patterns chart authors want to present to others.

\subsection{Approach Overview}

To achieve the above goals, we propose an approach to suggestions for improvements in chart design based on graphical patterns identified through a perception simulation model, as shown in Fig.~\ref{fig:pip}. Our approach starts with extracting visual effects from graphical elements, which serve as perceptual stimuli (\textbf{G1}) and input for the perception simulation model (Fig.~\ref{fig:pip} (b)). The extraction process takes charts in SVG files as input. We disassemble graphical elements according to SVG elements in files (Fig.~\ref{fig:pip} (a1)). Each element is characterized by a vector encoding its visual effects (see Fig.~\ref{fig:pip} (a2)), such as element types, adjacency relationships with other elements, position in the chart, filling color, etc. Next, a perception simulation model is then trained on human-annotated graphical patterns to learn how visual effects collectively form perceptually meaningful patterns (Fig.~\ref{fig:pip} (b1)). Once trained, the model can generate element representations (Fig.~\ref{fig:pip} (b2)) that enable perception-aware identification of element groups, in other words, graphical patterns (Fig.~\ref{fig:pip} (c1)). Also, the representation generation process yields quantifiable measures for evaluating the perceptual salience of element groups. Specifically, for each element group, we predict how it is perceived (\textbf{G2}) according to the commonality of intragroup elements and the differences with other elements (Fig.~\ref{fig:pip} (c2)). Chart authors can explore the perception effects of any element group (Fig.~\ref{fig:pip} (d1)), such as patterns identified by the perception simulation model or patterns expected by chart authors. Finally, we generate suggestions on improvements of chart design (Fig.~\ref{fig:pip} (d2), \textbf{G3}) based on the perceptual salience assessment of an expected pattern specified by chart authors. 

\subsection{Perception-based Pattern Identification}
\label{sec:dpm}
This section details the design of our proposed perception simulation model and demonstrates how it integrates perception grouping theories and how the model is applied to identify graphical patterns in charts.

\begin{figure}[htp]
    \centering
    \includegraphics[width=0.9\columnwidth]{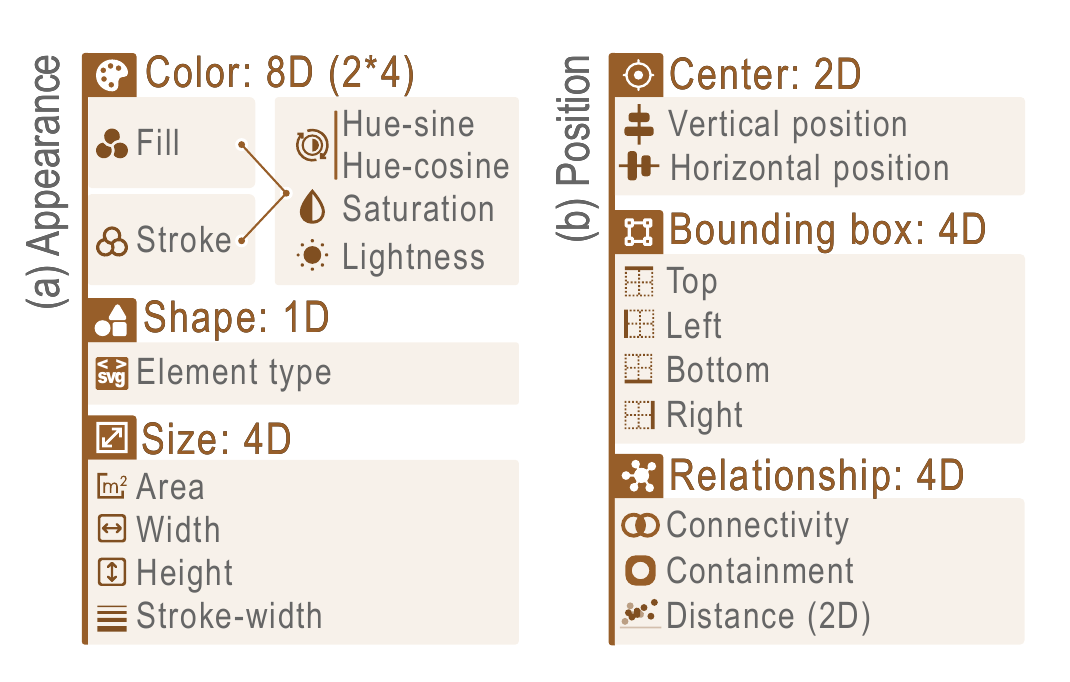}
    \caption{23-dimensional visual effects.}
    \label{fig:vf}
\end{figure}

\textbf{Characterizing visual effects.} For each element disassembled from the chart uploaded by chart authors, we record visual effects that could affect viewers' pattern perception results, as mentioned by Sec.~\ref{sec:ve}. As shown in Fig.~\ref{fig:vf} (a), we refine 13 dimensions of appearance effects. To represent color in a perceptually meaningful way, we convert all color attributes using the HSL (Hue, Saturation, Lightness) model. When opacity is part of the color encoding, the blended result on the canvas serves as the basis for HSL conversion. Additionally, the periodic nature of the hue, an angular dimension, is addressed by decomposing it into sine and cosine components, ensuring consistency with periodic representation. As for the shape dimension, we generate corresponding descriptions based on the type of SVG elements. The principle of proximity focuses on the position of elements. As shown in Fig.~\ref{fig:vf} (b), position effects can be characterized from three perspectives, which are the position of the centroid of each element, the bounding box coordinates, and inter-element relationships. We can calculate the effects of position dimensions from the first two perspectives based on SVG attributes directly. To maintain the representation consistency of relationship features with other visual effects, we compute relationship matrices to record pairwise relationships and apply multidimensional scaling (MDS) to generate descriptive dimensions.

\textbf{Recording perception grouping results.} To capture how people perceptually group visual elements into graphical patterns, we sampled 40 charts from the official examples of D3.js\footnote{D3: \url{https://d3js.org/}} and Highcharts\footnote{Highcharts: \url{https://www.highcharts.com/}}. (Details of the chart collection process can be found in a supplemental file.) Thirty-five annotators (aged 20–30) each labeled 10 randomly assigned charts via an interactive interface, identifying all perceivable patterns, marking the corresponding element groups, and rating them on (1) intra-group consistency and (2) boundary distinctness. Annotators received a nominal bonus, and in total,  we collected 1,273 pattern annotations (mean 2.78 patterns/chart and 10.18 unique annotations/chart). To ensure broad coverage, charts were chosen to span common visualization types and a range of visual complexities. Details on selection criteria, complexity metrics, and full annotation statistics can be found in the supplemental file.

\textbf{Extracting behavior patterns of human perception.} We identified two types of element correlations from annotated element groups to serve as training data for the perception simulation model. The first type is positive correlations, which exist between element pairs likely to be perceptually grouped together. We consider all pairs of elements within each group with a top rating of intra-group consistency to have positive correlations. In contrast, elements in the same chart but sharing no consistent group could be used to generate negative pairs. To focus on significant negative correlations, we select negative pairs by sampling elements from outside the group for each element within the annotated group. The number of sampling iterations, which defines how many negative correlations are identified per within-group element, is determined by the boundary distinctness rated for each element group. We calculate the sampling probability for each external element by a Gaussian kernel, which is inversely correlated with its distance to the elements within the group. We prioritize adjacent pairs because they could have significant negative correlations that override the principle of proximity.

\textbf{Grouping visual elements as graphical patterns.} We formulate a perception simulation model to learn perception-aware representations of visual elements from human annotations and subsequently identify graphical patterns in charts. The perception simulation model employs a contrastive learning framework, motivated by evidence that human perception inherently relies on comparative mechanisms~\cite{dehaene2003neural}. For training data, we use positive and negative sample pairs derived from two types of element correlations identified from human annotation (\textit{G1.1}). Once trained, the perception simulation model can take visual elements as input and transform visual elements into representations that indicate how human perception interprets the element according to its visual effects in 23 dimensions (see Fig.~\ref{fig:vf}). Next, we can identify element groups as graphical patterns by measuring the similarity of element representations. To simulate diverse perception mechanisms, we trim the element representation to yield distinct sub-representations, which correspond to perceptual results from diverse perspectives (\textit{G1.2}). Each sub-representation generates a distinct similarity metric, producing alternative grouping rules for elements. In addition to element representations, the perception simulation model also yields \textit{perceptual weight} $\boldsymbol{w}$ for each visual element, which demonstrates how each dimension of visual effects affects the perceptual result of the element (\textit{G2.1}).

\textbf{Measuring pattern salience.} We calculate a salience metric for each pattern to rank their recommendations during pattern exploration. The salience $S$ of an element group $E$ is calculated based on the consistency of elements within the group and the differences between elements inside and outside the group (\textit{G2.2}), as shown in the equations below. Unlike traditional single-channel models such as Itti's classic approach~\cite{itti2002model}, which focus on isolated features like color or orientation, our model considers multi-channel consistency and inter-group differences. 
\begin{align}
S_E &= \frac{\text{Avg}(C(i, j)|i\in E, j\in E, i\neq j)}{\text{Avg}(C(i, j),|i\in E, j\notin E)}
\end{align}
\begin{align}
C(i,j) &=\text{cos}(|\boldsymbol{w}_i|\cdot v_i,|\boldsymbol{w}_j|\cdot v_j)
\end{align}
We employ $C$ to measure the consistency of visual effects between two visual elements (i.e., $i$ and $j$). $\boldsymbol{w}_i$ refers to the perceptual weight for the visual effects of the element $i$; $v_i$ is its visual effect vector. Here, the perceptual weight is used to align with the consistency based on human perception. 
 
\textbf{Summarizing identified graphical patterns.} We further organize identified patterns through overlap analysis. If two identified patterns exhibit an element overlap exceeding 80\%, we extract the intersecting elements as a \textit{core pattern} and mark the two patterns as a similar pair. 

\textbf{Generating suggestions.} To support \textit{G3.2}, our approach enables chart authors to designate an expected graphical pattern by selecting a group of visual elements. For the selected group, we first compute its salience and then propose practical optimizations according to the result of the salience assessment. The salience of the group (see Eq. (1)) can indicate the necessity for optimizations. As for specific optimization suggestions, we first summarize the current usage of all visual effect dimensions to infer the existing design logic. To avoid disrupting existing visual encodings, we prioritize unused visual effect dimensions that will cause no conflict and can maximize the salience increase of the group. Our model identifies conflicts based on built-in rules, such as height and the top and the bottom of the bounding box not being used simultaneously. We also suggest reconsidering certain visual effects that may serve secondary purposes, such as enhancing aesthetic appeal, but could be prioritized to emphasize the selected pattern. We identify such visual effects according to the characteristic of low utilization (i.e., exhibit low diversity). After the dimension is determined, we test different effects through sampling methods. For instance, we test red, blue, and green as alternatives for visual effects of hue. When a value that can significantly enhance the salience of the currently selected element group is identified, we generate suggestions and provide them to chart authors.

  \begin{figure*}[!t]
    \centering
    \includegraphics[width=0.9\textwidth]{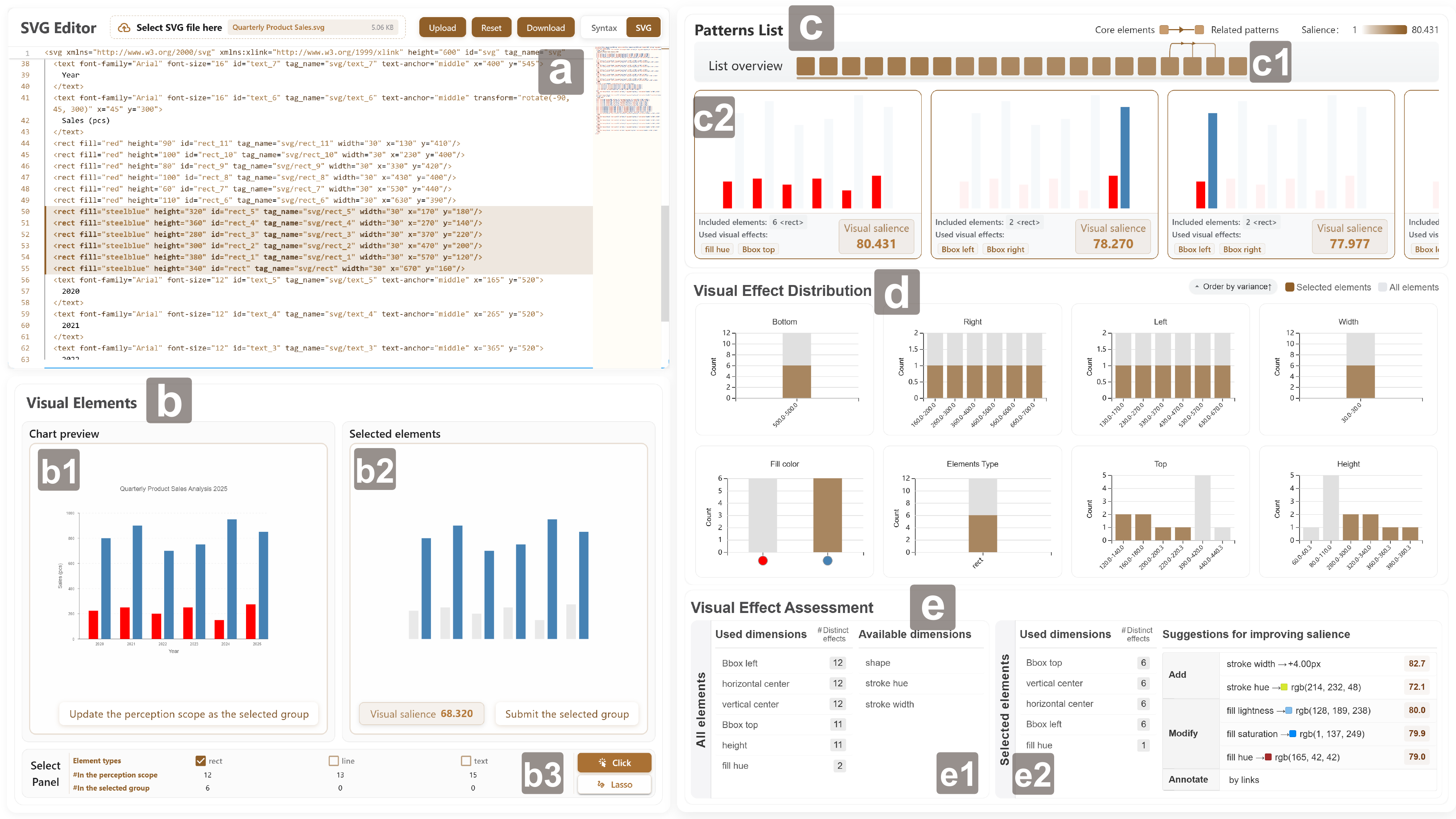}
  \caption{The interface of PatternSight for assessing the representation effectiveness of graphical patterns in charts. (a) The SVG Editor invites users to upload and edit chart files. (b) The visual elements view allows users to check the rendered chart and select element groups for pattern exploration. (c) The patterns list shows graphical patterns identified from the chart through human perception simulation. (d) The visual effect distribution summarizes the diversity of visual effects among the selected element group and the entire chart. (e) The visual effect assessment view describes the usage of visual effect dimensions and suggests chart optimizations that can increase the salience of the selected element group.}
    \label{fig:teaser}
  \end{figure*}

\subsection{Interface of PatternSight}
The interface of our tool consists of the SVG editor, the visual element view, the patterns list view, the visual effect distribution view, and the visual effect assessment view, as shown in Fig.~\ref{fig:teaser}. In this section, we introduce how the views coordinate to support our approach (Fig.~\ref{fig:pip}).

The SVG editor (Fig.~\ref{fig:teaser} (a)) allows chart authors to upload a chart by selecting an SVG file directly or providing chart-authoring source code that generates SVG based on commonly used generative syntax such as ECharts, Vega-Lite, D3, and Highcharts. To facilitate efficient design iteration, the editor incorporates online editing features such as reset, search, and replace.

The visual elements view (Fig.~\ref{fig:teaser} (b)) renders the uploaded chart to support the exploration of element groups. Before applying perception simulation, chart authors can remove irrelevant elements (e.g., legends) from the perception scope (Fig.~\ref{fig:teaser} (b1)) and select element groups of interest (Fig.~\ref{fig:teaser} (b2)) by click, lasso, and batch selection based on element types (Fig.~\ref{fig:teaser} (b3)). PatternSight calculates the salience score for the selected group of elements in real-time. More details of the group are provided in the visual effect distribution view (Fig.~\ref{fig:teaser} (d)) and the visual effect assessment view (Fig.~\ref{fig:teaser} (e2)).

The patterns list view (Fig.~\ref{fig:teaser} (c)) demonstrates graphical patterns identified by the perception simulation model through pattern cards (Fig.~\ref{fig:teaser} (c2)). In each card, we show the elements belonging to the pattern and list descriptions, consisting of 1) statistical results of element types, 2) dimensions of visual effects that contribute to the pattern identification, and 3) the salience score of the pattern. We sort the patterns by the salience score to prioritize prominent patterns. To further improve browsing efficiency, we provide a list overview (Fig.~\ref{fig:teaser} (c1)) at the top of the view. In the list overview, each pattern is presented by a square whose color encodes the salience of the pattern. By clicking a square, chart authors can jump to the corresponding pattern card. We also link squares representing patterns that have a large overlap of elements with their core pattern (please refer to Sec.\ref{sec:dpm}). According to the links, chart authors can identify and browse similar patterns quickly. 

The visual effect distribution view (Fig.~\ref{fig:teaser} (d)) utilizes histograms to display statistical results of activated dimensions of visual effects in the chart. When a group of elements is selected in the visual elements view or a pattern is chosen from the patterns list, the corresponding histogram highlights the selected group's proportion in brown. Besides, chart authors can figure out the usage of visual effect dimensions from different perspectives by sorting histograms in ascending or descending order according to the variance of visual effects within the selected group---dimensions with low variance help cluster elements into patterns, while those with high variance reveal internal details of the patterns.

The visual effect assessment view (Fig.~\ref{fig:teaser} (e)) lists activated dimensions of visual effects and suggestions for improving the salience of the selected element group. The view is split into two panels to provide an assessment for all elements (Fig.~\ref{fig:teaser} (e1)) and the selected elements (Fig.~\ref{fig:teaser} (e2)). In the left section of each panel, we count the number of distinct effects derived from each dimension in the corresponding element scope and sort dimensions according to the number. On the right section of the all elements panel, we list available dimensions of visual effects that are unused and whose inclusion does not cause visual clutter. Chart authors can prioritize available dimensions if they need to enrich the information represented by the chart. On the right section of the selected elements panel, we assess the salience score of the selected group and list improvement suggestions if the score is unsatisfactory. Suggestions are proposed from three perspectives, which are 1) adding a new dimension of visual effects to highlight the group through consistent visual effects, 2) modifying visual effects that contribute to internal inconsistency, and 3) adding additional annotations. PatternSight assists chart authors in adopting suggestions by recommending visual effects in detailed, code-based expressions. For instance, when suggesting ``set the stroke hue to 0,'' PatternSight provides additional code-based descriptions: ``stroke color $\rightarrow$ rgb(234, 0, 0).'' When multiple suggestions arise from the first two improvement perspectives, suggestions are ranked according to their predicted impact on increasing the salience score. If chart authors are interested in adding suggestions or modifying suggestions, they can apply the corresponding code editing by clicking the suggestion. PatternSight also supports value adjustment before applying. For annotation-related suggestions, we recommend either surrounding or connecting annotations, depending on whether the elements are adjacent. 

\begin{figure*}[htp]
    \centering
    \includegraphics[width=1.95\columnwidth]{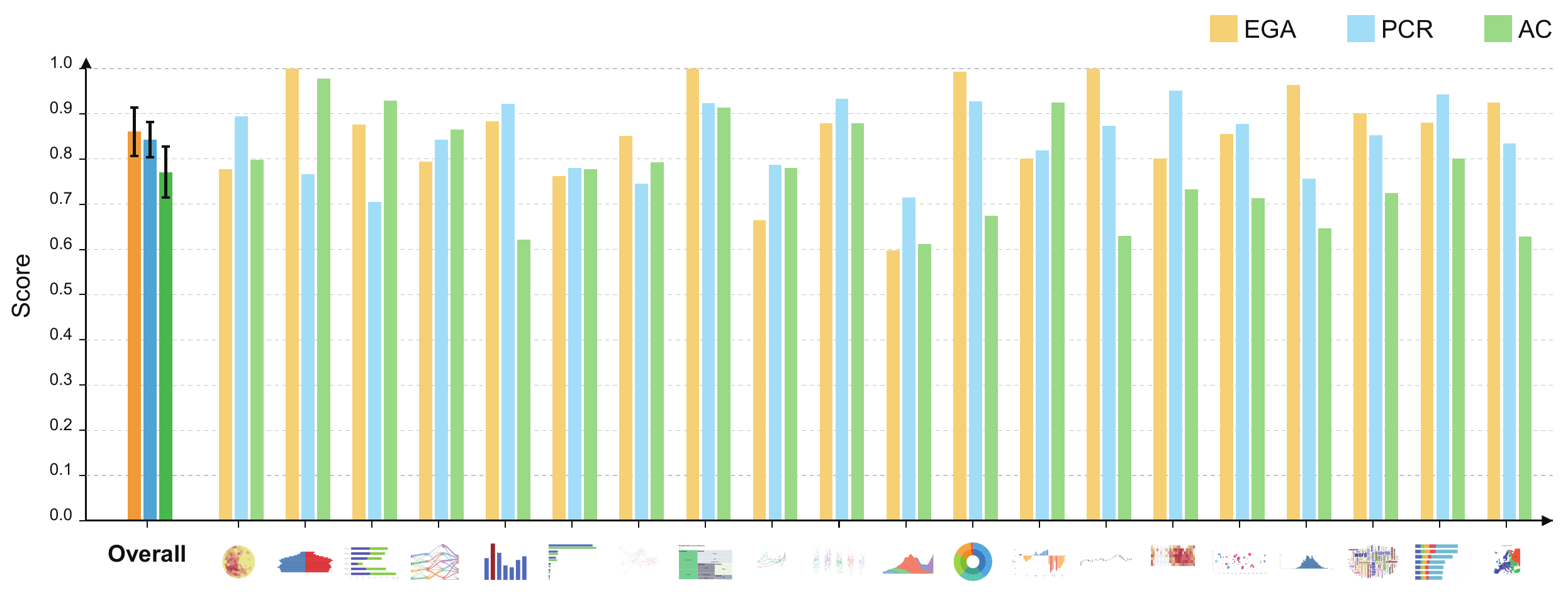}
    \caption{The assessment result of the perception simulation model. The error bars on the overall group represent a standard deviation of uncertainty.}
    \label{fig:qe}
\end{figure*}
\section{Evaluation}
We conducted a performance evaluation of PatternSight's perception simulation model and executed a user study to assess its interface. Two case studies were developed from the exploration processes implemented by two participants in the user study.

\subsection{Performance Evaluation}
We implemented a performance evaluation of the perception simulation model to validate its generalizability across diverse charts for pattern perception tasks. To generate a validation dataset, we replicated the annotation collection process for recording perception behaviors (see Sec.~\ref{sec:dpm}). This time, we invited 20 participants to annotate graphical patterns in 20 new charts (as shown on the horizontal axis of Fig.~\ref{fig:qe}). The validation dataset included 812 annotations of graphical patterns. Each chart received three annotations from each annotator on average.

\textbf{Metric design.} We designed three evaluation metrics to systematically compare the annotated results with the model output. The first two metrics focus on comparing the element membership between annotated patterns and element groups identified by the model, while the third metric compares the association rules for elements. To address the variability in the number of identified groups, we constructed equivalent comparisons by selecting the element groups with top-$k$ salience as the model output for the chart where $k$ graphical patterns were annotated collectively by all participants. 

1) Element group accuracy (EGA) measures the consistency between the model output and human perception (\textit{G1.1}) by calculating the average Jaccard similarity between each element group in the model output and the most similar graphical pattern annotated by participants to the group. 

2) Pattern coverage rate (PCR) verifies how the identified element groups can cover various patterns in human annotation (\textit{G1.2}) by calculating the average Jaccard similarity between each graphical pattern annotated by humans and the most similar element group identified by the model. 

3) Association consistency (AC) evaluates the alignment between the grouping rules employed in our model and the implicit element correlation principles underlying human perceptual mechanisms. Specifically, we generate an association matrix of elements based on the annotation results and the model, respectively. The third metric calculates the cosine similarity of the two matrices after normalization. Each cell of the matrix counts the times two elements are included in the same annotated pattern/identified element group.

\textbf{Result analysis.} As shown in Fig.~\ref{fig:qe}, we have separately assessed the perception simulation model based on its output for each chart and calculated the average score to get an overall assessment. We drew conclusions according to the overall assessment based on each metric: 1) EGA = 0.860: the output of the perception simulation model is highly consistent with human perceptual results; 2) PCR = 0.843: the model output can effectively cover the diversity of perceptual results; and 3) AC = 0.771: the perception simulation model can replicate the associative mechanisms of human perception regarding visual elements. Meanwhile, we observed that our model behaves unsatisfactorily in certain cases. For instance, our model received the lowest PCR (= 0.597) and lowest AC (= 0.611) in the simulation experiment for a stacked area chart, where individual elements exhibit irregular shapes. The boundary extraction mechanism employed in our model fails to accurately capture associative relationships among such elements. We discussed this issue in Sec.~\ref{sec:cc}.

\subsection{User Study}

To examine the practical utility and user experience of PatternSight in real-world visualization authoring scenarios, we conducted a within-subjects user study with 15 participants recruited from campus. All participants had prior experience using visualization tools such as Tableau, ECharts, or D3.js but lacked formal training in visual perception theory. The goal of the study was to evaluate the effectiveness of PatternSight in assisting chart authors with optimizing chart designs for graphical pattern representation. Each participant received a \$7 reward after the following procedure. 

\subsubsection{Procedure}
Before beginning the study, each participant attended a 25-minute tutorial that introduced the core functionalities of PatternSight. This session included a PowerPoint presentation explaining the theoretical background, followed by a live demonstration showcasing the tool’s capabilities in analyzing and modifying graphical patterns. Participants familiarized themselves with the interface and functionalities of PatternSight before proceeding to the main study tasks, consisting of the training phase and the exploration phase.

\textbf{Training phase.} Participants were assigned an SVG-based chart, a specified data pattern that is represented in the chart, and structured tasks, which consisted of adjusting visual elements to improve the perceptual grouping effectiveness of the specified pattern, and summarizing the improvement. They then used PatternSight to implement the tasks.

\textbf{Exploration phase.} In this phase, participants were given the freedom to interact with PatternSight without structured guidelines. They were encouraged to explore different visualizations, interact with PatternSight, experiment with optimizing graphical patterns, and use a think-aloud protocol. Throughout this phase, we documented participants' behavior, including their observations, interactions, and reflections on how the tool influenced their decision-making process.

\begin{figure}[ht]
    \centering
    \includegraphics[width=0.85\columnwidth]{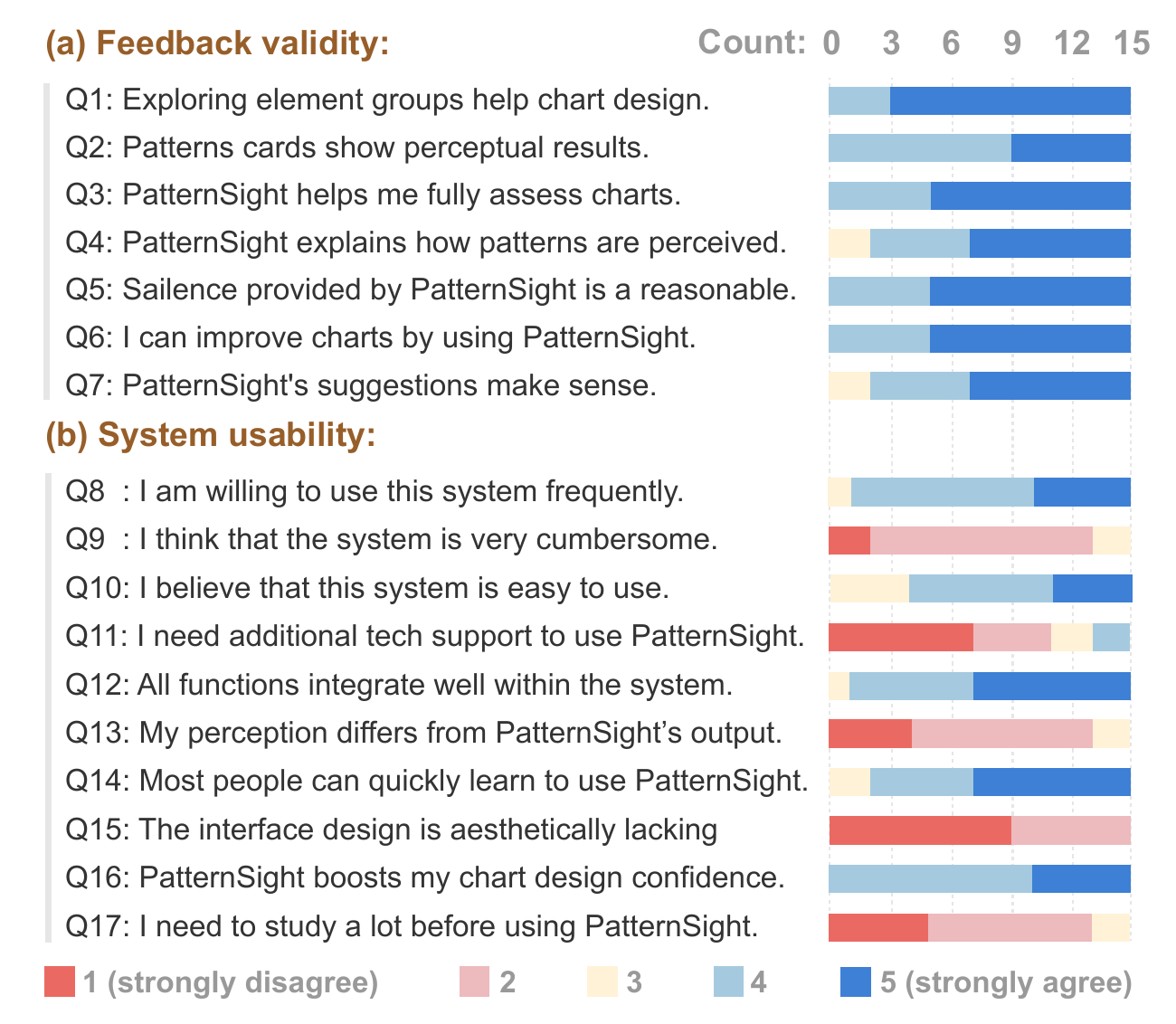}
    \caption{Quantitative results from questionnaire responses to: (a) seven items on feedback validity and (b) ten items for System Usability Scale.}
    \label{fig:ques}
\end{figure}

When all study tasks were finished, we collected user feedback through a questionnaire including two groups of quantitative questions using a 5-point Likert scale (see Fig.~\ref{fig:ques}) and open-ended questions for answer explanation.

\subsubsection{Results and Analysis}

All participants told us that they had encountered difficulties in enhancing graphical patterns in charts. The primary reported difficulty was that they could hardly decide which adjustments of visual design would be most effective, as they received no feedback or guidance on their design choices. The lack of immediate visual validation led to trial-and-error approaches, where they experimented with modifications but remained unsure whether their modifications enhanced the overall perceptual effectiveness of the visualization. As a result, the overall chart optimization process was time-consuming and inefficient. In contrast, PatternSight demonstrated significant advantages in assisting participants with graphical pattern identification and modification. We discuss participants' feedback from the following two aspects.

\textbf{Feedback validity (Q1-Q7).} Our questionnaire included questions probing the validity of feedback provided by PatternSight in optimizing chart design and facilitating an understanding of perceptual outcomes. Participants consistently highlighted PatternSight’s ability to streamline chart design, particularly through its graphical patterns list. One participant noted, ``The list of graphical patterns greatly assists in selection and is easy to operate.'' Furthermore, PatternSight’s capacity to explain the perceptual mechanism of graphical patterns was frequently praised, with at least two participants emphasizing its educational value. For instance, a participant remarked, ``PatternSight helps me understand how a pattern is perceived.'' PatternSight can not only aid chart authors in design optimization but also enhance their awareness of perceptual theories, enabling them to identify previously unnoticed patterns, and finally leading to a better chart design experience. We also received suggestions, such as ``optimization suggestions could be provided from more perspectives.'' Based on the issues encountered by participants during the exploration phase, we identified modifying scale encodings as future work.

\textbf{System usability (Q8-Q17).} Our questionnaire collected feedback from aspects, including ease of usage, interface aesthetics, and functional integration. Overall, PatternSight received a usability score of 81.2. Most participants agreed that PatternSight can directly address practical needs in chart authoring, such as identifying groups of elements, getting suggestions on chart optimization, and testing modifications. For example, a participant commented, ``Clicking on the suggestion can be directly applied, and the code is automatically modified, greatly reducing the time cost.'' Additionally, two participants suggested improving the element selection experience, especially when dealing with charts with high element density or charts containing extremely small elements.

\subsection{Case1: Salience Balance between Groups}
\label{sec:c1}

Our participant, using Alice as an alias, was instructed to assume the role of a revenue analyst who needed to present an initial visualization depicting six-year sales data for two product categories (A and B) using a bar chart. As shown in Fig.~\ref{fig:teaser} (b1), the chart encoded sales values through bar height while employing color to distinguish product categories--red for category A and steel blue for category B. Alice aimed to achieve comparable salience levels across graphical patterns representing each product category, as she sought to avoid introducing categorical bias in this data presentation. After the chart was loaded, the patterns identified by PatternSight indicated that the bars for each category and each pair of adjacent bars could be perceived as graphical patterns. As shown in the first card in Fig.~\ref{fig:teaser} (c2), the pattern consisting exclusively of red bars emerged as the most salient pattern in the chart. PatternSight identified ``fill hue'' and ``Bbox top'' as the main factors contributing to the high salience---these bars featured uniformly vibrant red fills and shared similarly low vertical positions of their top bounding boxes. However, the pattern of blue bars (i.e., the other product category) had a relatively low salience, even lower than several patterns composed of adjacent bar pairs with color variations, as shown in Fig.~\ref{fig:teaser} (c2). Alice inferred that the differences in color hue and lightness significantly influenced salience.

\begin{figure}[htp]
\centering
\includegraphics[width=0.75\columnwidth]{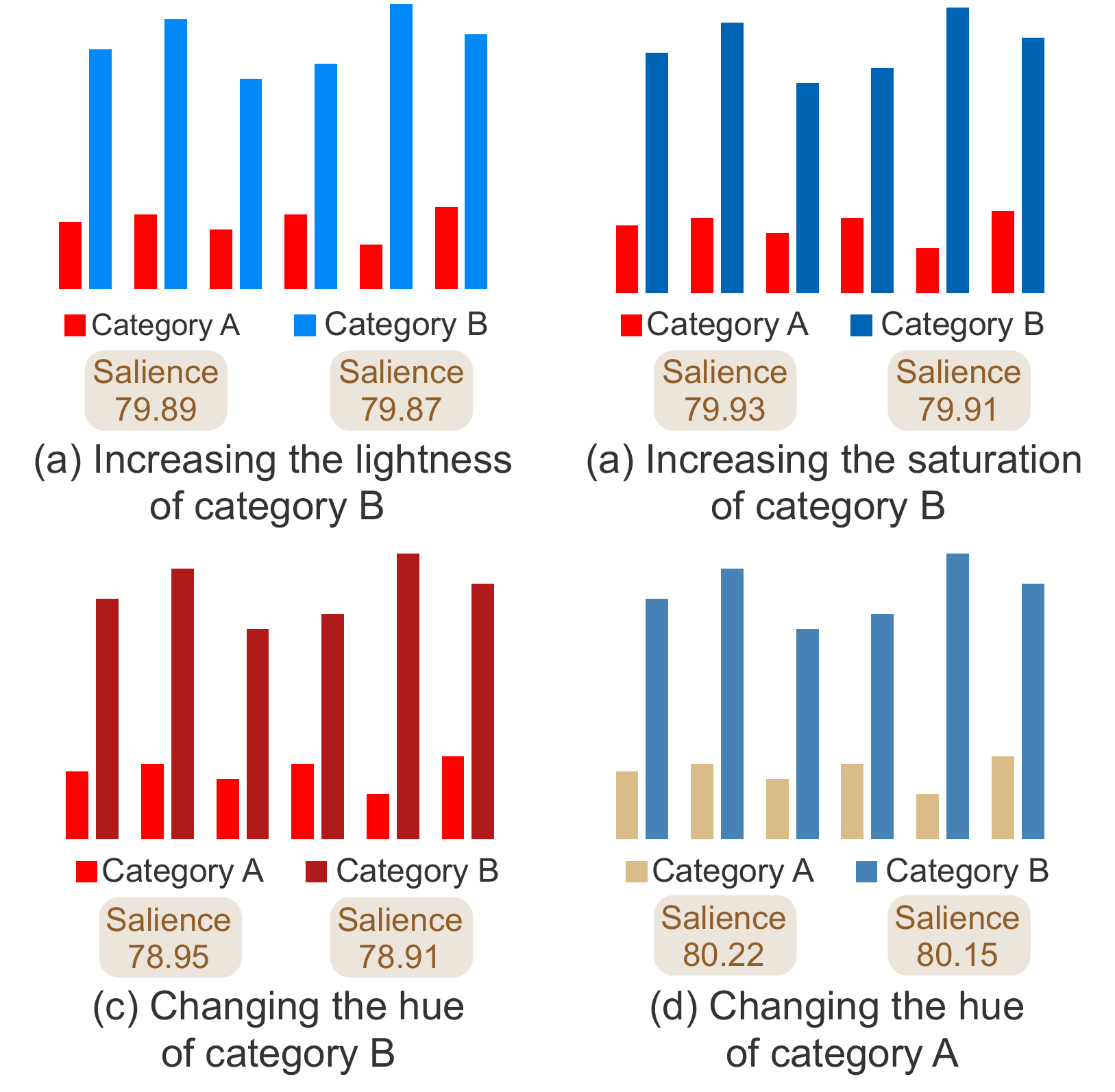}
\caption{Four feasible optimization schemes to balance the pattern salience.}
\label{fig:c1}
\end{figure}

To balance the salience levels of the two product categories, Alice clicked the pattern card of category B to select all blue bars as the element group for salience improvement. Next, Alice browsed suggestions in the visual effect assessment view (Fig.~\ref{fig:teaser} (e2)), which included adding stroke effects, modifying fill color, and adding annotations of links. Alice preferred the scheme of modifying the fill color because the other two would compromise the chart's intuitive readability. Specifically, the fill color could be changed by increasing the lightness (Fig.~\ref{fig:c1} (a)) or the saturation of the blue (Fig.~\ref{fig:c1} (b)) or setting the hue to dark red (Fig.~\ref{fig:c1} (c)). Alice employed the batch modification feature to test the three schemes. Although all three schemes can increase the salience rank of product category B, none of the suggested colors look in harmony next to the red bars. Alice then came up with a solution by reverse thinking---if it is unfeasible to improve the salience rank of the second group, we can also reduce the salience rank of the first group to achieve salience balance. Alice determined to modify the fill color for product category A. Alice considered that red was inappropriate for the bars of category A because the red color appears to sarcastically mock the low sales. Alice preferred to present the revenue data with a positive mindset. Brown would be a good choice. After several attempts, Alice finally selected light brown as the color effect for Category A, as shown in Fig.~\ref{fig:c1} (d). The salience metric for the two groups measured 80.22 and 80.15 respectively, demonstrating comparable visual prominence while maintaining aesthetic coherence.

\subsection{Case2: Highlights on Habitable Exoplanets}

\begin{figure}[!htb]
\centering
\includegraphics[width=0.65\columnwidth]{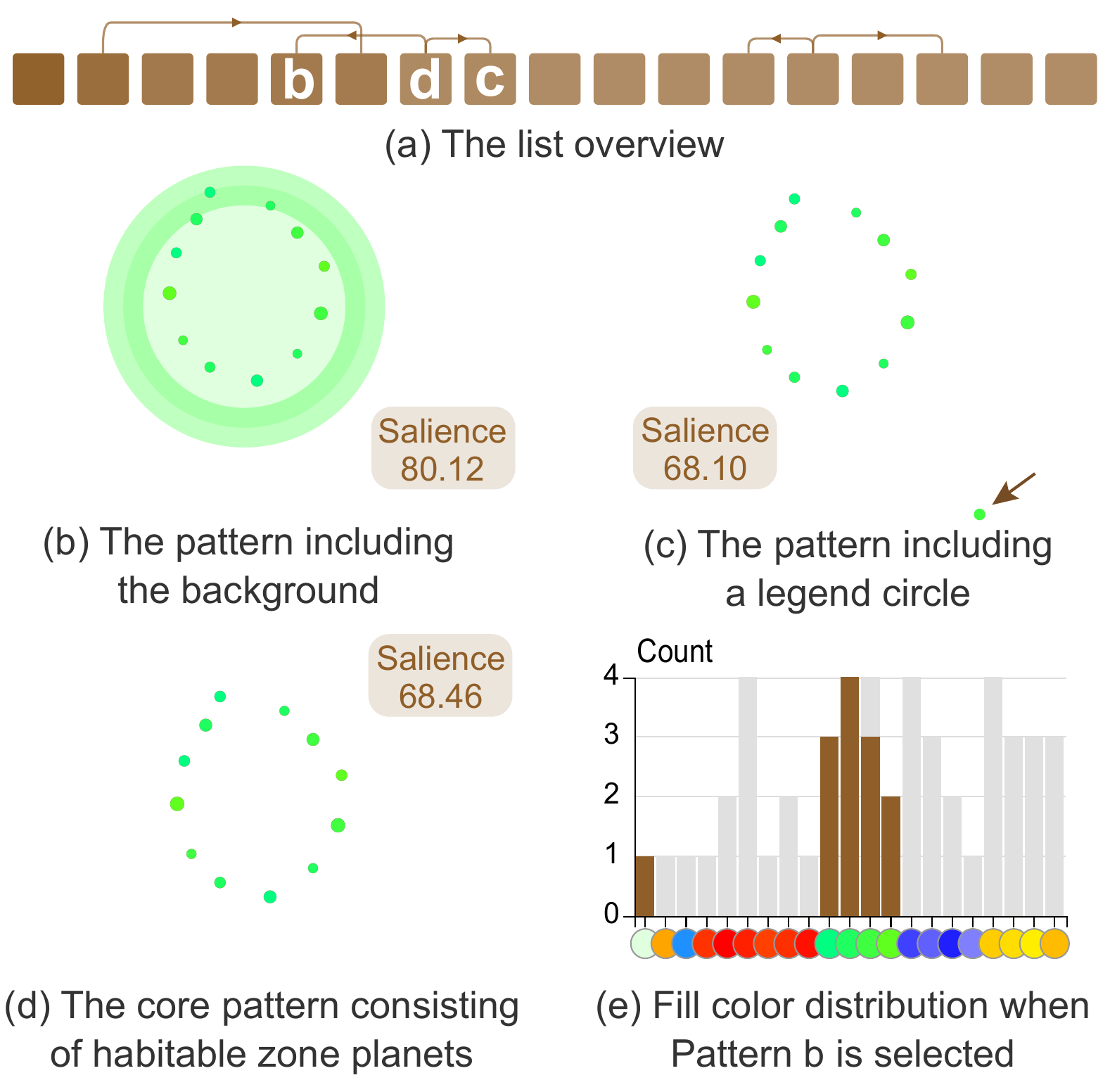}
\caption{Three similar patterns identified by PatternSight.}
\label{fig:c2}
\end{figure}

In this case, the participant, using Bob as an alias, wanted to optimize a visualization (Fig.~\ref{fig:c2} (a)) of the Exoplanet Discovery Map that would be presented in a public-friendly astronomical exhibition. The initial visualization demonstrated over 60 exoplanets discovered between 2010 and 2023. Each exoplanet was shown as a circle whose polar coordinates encoded distance from the center represented orbital distance (in astronomical units) by radius and encoded discovery year by angle. The size of each circle is scaled to the planet’s mass relative to Earth. The color of each circle reflected the type of exoplanets, which could be red for large gas giants, green for habitable zone planets, blue for recently discovered, and yellow for others. The habitable zone planets could be the most interesting to viewers. Thus, Bob needed to improve the salience of green circles.

\begin{figure}[!htb]
\centering
\includegraphics[width=\columnwidth]{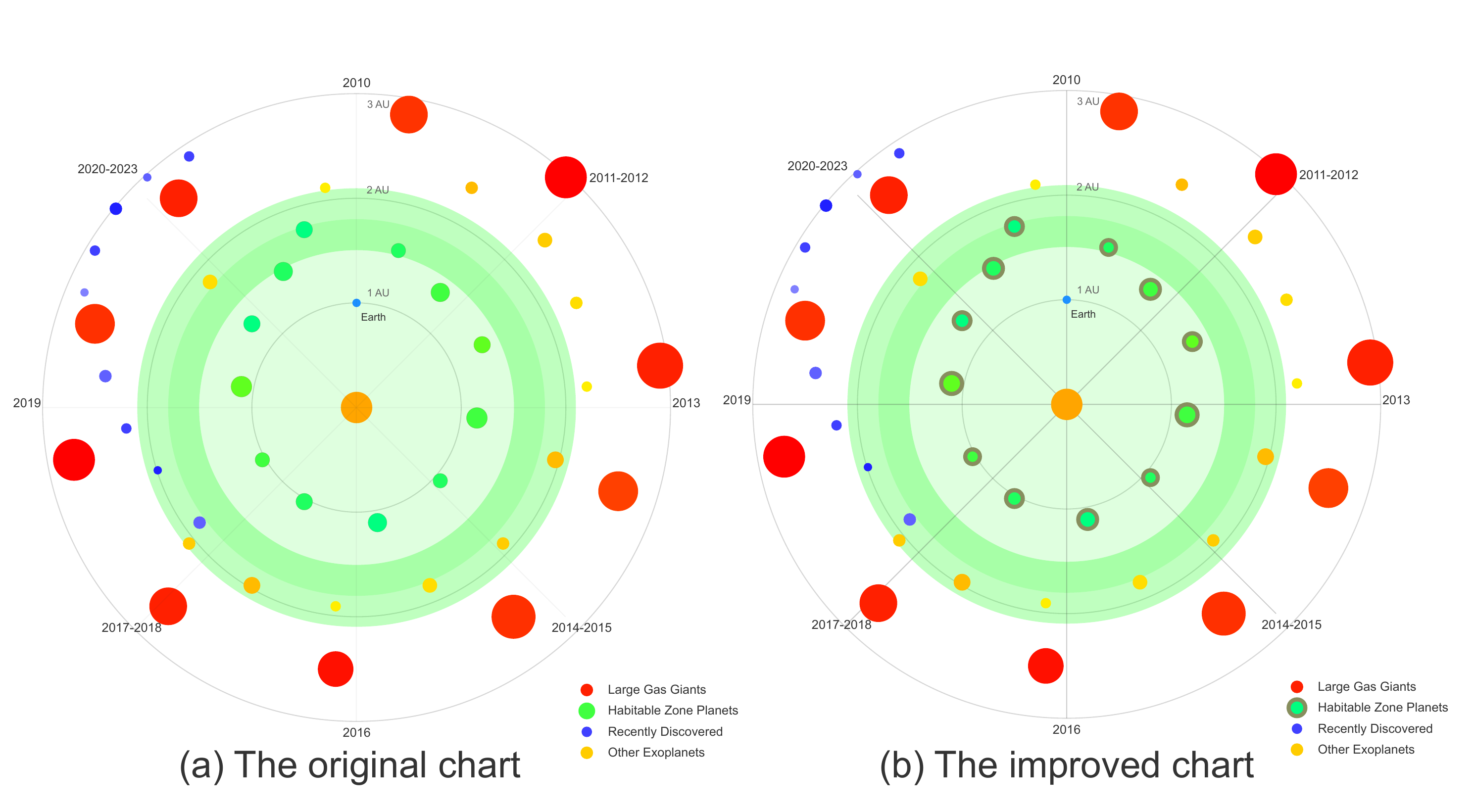}
\caption{The Exoplanet Discovery Map before and after chart edit using PatternSight.}
\label{fig:ca2}
\end{figure}

After loading the SVG file of the chart into PatternSight, Bob first filtered circle elements as the scope of perception. The identification result indicated that circles were grouped into multiple patterns by color, i.e., planet types. According to the patterns list view, Gas giants (large red dots) scored a striking 99.201 in salience. However, Bob was more concerned about the pattern formed by the green circles representing habitable planets. Scrolling backward through the pattern list, he found that the first such pattern included both the background circle for the habitable zone and the habitable zone planets. When this pattern was selected, the distribution of fill colors in the visual effect distribution view, as shown in Fig.~\ref{fig:c2} (e), revealed that there was another green element in the chart. The associated annotations in the list overview, as shown in Fig.~\ref{fig:c2} (a), led Bob to discover a pattern composed only of small green circles, but with an additional circle serving as the legend for a habitable planet (see Fig.~\ref{fig:c2} (c)). The salience score of this pattern was 68.10, lower than that of the pattern with the background (score: 80.12). Meanwhile, the core element group shared by both patterns—containing only the habitable planet elements (Fig.~\ref{fig:c2} (d))—also had a relatively lower salience score of 68.46. Bob believed that the large-area background could highlight the scope of the habitable planets but weakened the salience of the planets themselves.

He continued with the element group without the legend circle to check optimization suggestions. The suggestion panel lists no solution that could improve salience by adding visual effect dimensions because the visualization already employed many visual encodings. As for modifications, PatternSight recommended increasing the stroke width to 3 pixels and changing the stroke color to dark green. Fortunately, almost all circles were rendered with a default black stroke with a width of 0.1 pixel---the stroke effects were not used to represent any attributes. Bob thought that it was a good idea to take advantage of the stroke effects.

After adopting suggestions on stroke, the pattern salience of habitable planets rose from 68.41 to 80.08. Bob found that the bold dark green strokes looked like halos, as shown in Fig.~\ref{fig:ca2} (b). Although the actual size of the green circles had not changed, the outlines of those visual elements were amplified to highlight the green circles. He was satisfied with the new chart. He finally updated the visual effect of the legend circle in the SVG Editor and finished the chart optimization.


\section{Discussion}

\subsection{Contrastive Learning v.s. Computer Vision}
\label{sec:cc}
PatternSight employs a contrastive learning framework to identify graphical patterns from charts by simulating the relativity of human visual perception. After integrating perceptual grouping theory, our model can achieve satisfactory results after training with a small amount of human annotation. Most recently proposed image recognition techniques are mainly based on CV. In particular, large-scale vision models demonstrate extraordinary performance in recognizing and understanding image semantics based on training data or prompts. However, the mechanisms by which human vision captures graphical patterns have not yet generated a sufficient volume of data to support relevant training tasks. Additionally, vision models, with their vast number of parameters and complex architectures, produce outputs whose mechanisms are as difficult to interpret as perceptual mechanisms, which hinders chart authors from instilling confidence in suggested chart improvement proposals from those models. 

By incorporating perceptual grouping theory, our model achieves accurate perceptual simulation with few human annotations while explaining how visual effects drive perceptual grouping. However, our model struggles with complex boundaries (e.g., irregular shapes) due to the simplified understanding of boundaries based on bounding boxes. Future improvements will integrate CV models to better describe irregular shapes and spatial relationships.

\subsection{Suggestions for Scale Encodings}
\label{sec:sug}
PatternSight provides suggestions on chart optimization from the perspective of the visual effects of elements. According to the usage experience recorded in the user study (e.g., the two cases), such suggestions are feasible and effective for charts. However, there could be requirements to get suggestions from the perspective of visual encoding. For example, when representing numerical attributes, chart authors prefer scaling encodings, like gradient-based color or a numerical axis. Currently, PatternSight can only suggest improvements in scaling encodings by batch editing, such as translating the entire horizontal position by 30 pixels or changing the hue of a color gradient with an intensity scale to green. SVG files contain only visual rendering information, not the underlying data being visualized. Thus, it is challenging to extract precise mapping mechanisms from user-uploaded SVG files, consequently limiting our ability to provide relevant suggestions. We plan to address this limitation in future work by understanding the chart authoring code.

\subsection{From Assistance Tool to Cognitive Scaffold} 
\label{sec:cognitive}
Unlike existing visualization tools that focus on generating new views or handling layout adjustments, our approach evaluates the perceptual grouping effectiveness of existing designs. The key insight from our user study is that PatternSight's value lies not only in chart optimization schemes but also in the explanation of perceptual grouping. PatternSight can function as a ``cognitive scaffold'' by correlating complex design goals for perceptual grouping with intuitive design variables. For example, as mentioned in Sec.~\ref{sec:c1}, Alice evolved from vague goals of ``balancing visual appeal'' to achieving visual harmony through informed color adjustments. 


\section{Conclusion}
Chart authors who lack formal training in perceptual theory may face challenges in evaluating the effectiveness of their designs and selecting optimal representations to convey data patterns. To address the challenge, we proposed a perception simulation model that can predict observable graphical patterns, thereby assessing a chart’s perceptual effectiveness. Based on the relativity of visual perception, we selected contrastive learning as the framework of our model and fully considered the perceptual grouping theories in the feature selection process. Empirical validation confirmed that the model effectively simulates human perceptual behaviors, approximating how real viewers interpret charts.

To apply the model, we developed PatternSight\footnote{\url{https://github.com/729149195/PatternSight}}, an interactive tool that enables chart authors to evaluate whether their designs effectively emphasize intended data patterns and provide feasible suggestions for visual refinements. The result of the user study demonstrated that PatternSight can spark more ideas on chart optimization and bolster the confidence of chart authors during the chart refinement process.

\bibliography{template}

\begin{thebibliography}{10}
\providecommand{\url}[1]{#1}
\csname url@samestyle\endcsname
\providecommand{\newblock}{\relax}
\providecommand{\bibinfo}[2]{#2}
\providecommand{\BIBentrySTDinterwordspacing}{\spaceskip=0pt\relax}
\providecommand{\BIBentryALTinterwordstretchfactor}{4}
\providecommand{\BIBentryALTinterwordspacing}{\spaceskip=\fontdimen2\font plus
\BIBentryALTinterwordstretchfactor\fontdimen3\font minus \fontdimen4\font\relax}
\providecommand{\BIBforeignlanguage}[2]{{%
\expandafter\ifx\csname l@#1\endcsname\relax
\typeout{** WARNING: IEEEtran.bst: No hyphenation pattern has been}%
\typeout{** loaded for the language `#1'. Using the pattern for}%
\typeout{** the default language instead.}%
\else
\language=\csname l@#1\endcsname
\fi
#2}}
\providecommand{\BIBdecl}{\relax}
\BIBdecl

\bibitem{wagemans2012century}
J.~Wagemans, J.~H. Elder, M.~Kubovy \emph{et~al.}, ``{A Century of Gestalt Psychology in Visual Perception: I. Perceptual grouping and figure--ground organization.}'' \emph{Psychological bulletin}, vol. 138, no.~6, p. 1172, 2012.

\bibitem{wongsuphasawat2017voyager}
K.~Wongsuphasawat, Z.~Qu, D.~Moritz \emph{et~al.}, ``{Voyager 2: Augmenting Visual Analysis with Partial View Specifications},'' in \emph{Proceedings of the 2017 CHI Conference on Human Factors in Computing Systems}, pp. 2648--2659.

\bibitem{wang2022towards}
Y.~Wang, Z.~Hou, L.~Shen \emph{et~al.}, ``Towards natural language-based visualization authoring,'' \emph{IEEE Transactions on Visualization and Computer Graphics}, vol.~29, no.~1, pp. 1222--1232, 2022.

\bibitem{chen2019marvist}
Z.~Chen, Y.~Su, Y.~Wang \emph{et~al.}, ``{MARVisT: Authoring Glyph-Based Visualization in Mobile Augmented Reality},'' \emph{IEEE Transactions on Visualization and Computer Graphics}, vol.~26, no.~8, pp. 2645--2658, 2019.

\bibitem{liu2018data}
Z.~Liu, J.~Thompson, A.~Wilson \emph{et~al.}, ``{Data Illustrator: Augmenting Vector Design Tools with Lazy Data Binding for Expressive Visualization Authoring},'' in \emph{Proceedings of the 2018 CHI Conference on Human Factors in Computing Systems}, pp. 1--13.

\bibitem{wang2021falx}
C.~Wang, Y.~Feng, R.~Bodik \emph{et~al.}, ``{Falx: Synthesis-Powered Visualization Authoring},'' in \emph{Proceedings of the 2021 CHI Conference on Human Factors in Computing Systems}, 2021, pp. 1--15.

\bibitem{rubab2021examining}
S.~Rubab, J.~Tang, and Y.~Wu, ``Examining interaction techniques in data visualization authoring tools from the perspective of goals and human cognition: a survey,'' \emph{Journal of Visualization}, vol.~24, pp. 397--418, 2021.

\bibitem{quadri2024you}
G.~J. Quadri, A.~Z. Wang, Z.~Wang \emph{et~al.}, ``{Do You See What I See? A Qualitative Study Eliciting High-Level Visualization Comprehension},'' in \emph{Proceedings of the 2024 CHI Conference on Human Factors in Computing Systems}, pp. 1--26.

\bibitem{munzner2009nested}
T.~Munzner, ``{A Nested Model for Visualization Design and Validation},'' \emph{IEEE Transactions on Visualization and Computer Graphics}, vol.~15, no.~6, pp. 921--928, 2009.

\bibitem{bresciani2015pitfalls}
S.~Bresciani and M.~J. Eppler, ``{The Pitfalls of Visual Representations: A Review and Classification of Common Errors Made While Designing and Interpreting Visualizations},'' \emph{Sage Open}, vol.~5, no.~4, pp. 1--14, 2015.

\bibitem{qu2017keeping}
Z.~Qu and J.~Hullman, ``{Keeping Multiple Views Consistent: Constraints, Validations, and Exceptions in Visualization Authoring},'' \emph{IEEE Transactions on Visualization and Computer Graphics}, vol.~24, no.~1, pp. 468--477, 2018.

\bibitem{lo2022misinformed}
L.~Y.-H. Lo, A.~Gupta, K.~Shigyo \emph{et~al.}, ``{Misinformed by Visualization: What Do We Learn From Misinformative Visualizations?}'' \emph{Computer Graphics Forum}, vol.~41, no.~3, pp. 515--525, 2022.

\bibitem{lam2011empirical}
H.~Lam, E.~Bertini, P.~Isenberg \emph{et~al.}, ``{Empirical Studies in Information Visualization: Seven Scenarios},'' \emph{IEEE Transactions on Visualization and Computer Graphics}, vol.~18, no.~9, pp. 1520--1536, 2011.

\bibitem{elliott2020design}
M.~A. Elliott, C.~Nothelfer, C.~Xiong, and D.~A. Szafir, ``{A Design Space of Vision Science Methods for Visualization Research},'' \emph{IEEE Transactions on Visualization and Computer Graphics}, vol.~27, no.~2, pp. 1117--1127, 2021.

\bibitem{quadri2021survey}
G.~J. Quadri and P.~Rosen, ``{A Survey of Perception-Based Visualization Studies by Task},'' \emph{IEEE Transactions on Visualization and Computer Graphics}, vol.~28, no.~12, pp. 5026--5048, 2022.

\bibitem{gramazio2014relation}
C.~C. Gramazio, K.~B. Schloss, and D.~H. Laidlaw, ``The relation between visualization size, grouping, and user performance,'' \emph{IEEE Transactions on Visualization and Computer Graphics}, vol.~20, no.~12, pp. 1953--1962, 2014.

\bibitem{satyanarayan2019critical}
A.~Satyanarayan, B.~Lee, D.~Ren \emph{et~al.}, ``{Critical Reflections on Visualization Authoring Systems},'' \emph{IEEE Transactions on Visualization and Computer Graphics}, vol.~26, no.~1, pp. 461--471, 2020.

\bibitem{bako2022streamlining}
H.~Bako, A.~Varma, A.~Faboro \emph{et~al.}, ``{Streamlining Visualization Authoring in D3 Through User-Driven Templates},'' in \emph{Proceedings of 2022 IEEE Visualization and Visual Analytics}, pp. 16--20.

\bibitem{way2018unconventional}
R.~I.~T. Way, ``for unconventional statistical graphs,'' 2018.

\bibitem{wootton2024charting}
D.~Wootton, A.~R. Fox, E.~Peck, and A.~Satyanarayan, ``Charting eda: Characterizing interactive visualization use in computational notebooks with a mixed-methods formalism,'' \emph{IEEE Transactions on Visualization and Computer Graphics}, 2024.

\bibitem{sultanum2021leveraging}
N.~Sultanum, F.~Chevalier, Z.~Bylinskii, and Z.~Liu, ``{Leveraging Text-Chart Links to Support Authoring of Data-Driven Articles with VizFlow},'' in \emph{Proceedings of the 2021 CHI Conference on Human Factors in Computing Systems}, pp. 1--17.

\bibitem{stoiber2023authoring}
C.~Stoiber, S.~Radkohl, F.~Grassinger \emph{et~al.}, ``{Authoring Tool for Data Journalists Integrating Self-Explanatory Visualization Onboarding Concept for a Treemap Visualization},'' in \emph{Proceedings of the 15th Biannual Conference of the Italian SIGCHI Chapter}, 2023, pp. 1--14.

\bibitem{wu2021multivision}
A.~Wu, Y.~Wang, M.~Zhou \emph{et~al.}, ``{MultiVision: Designing Analytical Dashboards with Deep Learning Based Recommendation},'' \emph{IEEE Transactions on Visualization and Computer Graphics}, vol.~28, no.~1, pp. 162--172, 2022.

\bibitem{kim2023dupo}
H.~Kim, R.~Rossi, J.~Hullman, and J.~Hoffswell, ``{Dupo: A Mixed-Initiative Authoring Tool for Responsive Visualization},'' \emph{IEEE Transactions on Visualization and Computer Graphics}, vol.~30, no.~1, pp. 934--943, 2024.

\bibitem{gupta1997visual}
A.~Gupta and R.~Jain, ``Visual information retrieval,'' \emph{Communications of the ACM}, vol.~40, no.~5, pp. 70--79, 1997.

\bibitem{shah2019kvqa}
S.~Shah, A.~Mishra, N.~Yadati, and P.~P. Talukdar, ``{KVQA: Knowledge-Aware Visual Question Answering},'' in \emph{Proceedings of the AAAI conference on artificial intelligence}, vol.~33, no.~01, 2019, pp. 8876--8884.

\bibitem{qiao2018exploring}
T.~Qiao, J.~Dong, and D.~Xu, ``{Exploring Human-Like Attention Supervision in Visual Question Answering},'' in \emph{Proceedings of the AAAI Conference on Artificial Intelligence}, vol.~32, no.~1, 2018.

\bibitem{kafle2020answering}
K.~Kafle, R.~Shrestha, S.~Cohen \emph{et~al.}, ``{Answering Questions about Data Visualizations using Efficient Bimodal Fusion},'' in \emph{Proceedings of the IEEE/CVF Winter conference on applications of computer vision}, 2020, pp. 1498--1507.

\bibitem{masry2022chartqa}
A.~Masry, X.~L. Do, J.~Q. Tan \emph{et~al.}, ``{ChartQA: A Benchmark for Question Answering about Charts with Visual and Logical Reasoning},'' in \emph{Proceedings of the Findings of the Association for Computational Linguistics}, 2022, pp. 2263--2279.

\bibitem{li2022structure}
H.~Li, Y.~Wang, A.~Wu \emph{et~al.}, ``{Structure-aware Visualization Retrieval},'' in \emph{Proceedings of the 2022 CHI Conference on Human Factors in Computing Systems}, pp. 1--14.

\bibitem{yi2008understanding}
J.~S. Yi, Y.-a. Kang, J.~T. Stasko, and J.~A. Jacko, ``{Understanding and Characterizing insights: How Do People Gain insights Using information Visualization?}'' in \emph{Proceedings of the 2008 Workshop on BEyond time and errors: novel evaLuation methods for Information Visualization}, 2008, pp. 1--6.

\bibitem{rosli2015gestalt}
M.~H.~W. Rosli and A.~Cabrera, ``{Gestalt Principles in Multimodal Data Representation},'' \emph{IEEE Computer Graphics and Applications}, vol.~35, no.~2, pp. 80--87, 2015.

\bibitem{ventocilla2020comparative}
E.~Ventocilla and M.~Riveiro, ``A comparative user study of visualization techniques for cluster analysis of multidimensional data sets,'' \emph{Information Visualization}, vol.~19, no.~4, pp. 318--338, 2020.

\bibitem{fox2023theories}
A.~R. Fox, ``Theories and models in graph comprehension,'' \emph{Visualization Psychology}, pp. 39--64, 2023.

\bibitem{fujiwara2019incremental}
T.~Fujiwara, J.-K. Chou, S.~Shilpika \emph{et~al.}, ``{An Incremental Dimensionality Reduction Method for Visualizing Streaming Multidimensional Data},'' \emph{IEEE Transactions on Visualization and Computer Graphics}, vol.~26, no.~1, pp. 418--428, 2020.

\bibitem{gramazio2016colorgorical}
C.~C. Gramazio, D.~H. Laidlaw, and K.~B. Schloss, ``{Colorgorical: Creating discriminable and preferable color palettes for information visualization},'' \emph{IEEE Transactions on Visualization and Computer Graphics}, vol.~23, no.~1, pp. 521--530, 2017.

\bibitem{silva2011using}
S.~Silva, B.~S. Santos, and J.~Madeira, ``Using color in visualization: A survey,'' \emph{Computers \& Graphics}, vol.~35, no.~2, pp. 320--333, 2011.

\bibitem{zeng2023review}
Z.~Zeng and L.~Battle, ``{A Review and Collation of Graphical Perception Knowledge for Visualization Recommendation},'' in \emph{Proceedings of the 2023 CHI Conference on Human Factors in Computing Systems}, 2023, pp. 1--16.

\bibitem{zeng2023too}
Z.~Zeng, J.~Yang, D.~Moritz \emph{et~al.}, ``{Too Many Cooks: Exploring How Graphical Perception Studies Influence Visualization Recommendations in Draco},'' \emph{IEEE Transactions on Visualization and Computer Graphics}, vol.~30, no.~1, pp. 1063--1073, 2023.

\bibitem{izakson2020proximity}
L.~Izakson, Y.~Zeevi, and D.~J. Levy, ``Attraction to similar options: The gestalt law of proximity is related to the attraction effect,'' \emph{PLoS ONE}, vol.~15, no.~10, p. e0240937, 2020.

\bibitem{pinna2022similarity}
B.~Pinna, D.~Porcheddu, and J.~Skilters, ``Similarity and dissimilarity in perceptual organization: On the complexity of the gestalt principle of similarity,'' \emph{Vision}, vol.~6, no.~3, p.~39, 2022.

\bibitem{palmer1994rethinking}
S.~Palmer and I.~Rock, ``Rethinking perceptual organization: The role of uniform connectedness,'' \emph{Psychonomic bulletin \& review}, vol.~1, no.~1, pp. 29--55, 1994.

\bibitem{palmer1992common}
S.~E. Palmer, ``Common region: A new principle of perceptual grouping,'' \emph{Cognitive psychology}, vol.~24, no.~3, pp. 436--447, 1992.

\bibitem{reeder2017individual}
R.~R. Reeder, ``Individual differences shape the content of visual representations,'' \emph{Vision Research}, vol. 141, pp. 266--281, 2017.

\bibitem{xu2024insights}
K.~Xu, ``Insights into the relationship between eye movements and personality traits in restricted visual fields,'' \emph{Scientific Reports}, vol.~14, no.~1, p. 10261, 2024.

\bibitem{bearfield2024same}
C.~X. Bearfield, L.~Van~Weelden, A.~Waytz, and S.~Franconeri, ``{Same Data, Diverging Perspectives: The Power of Visualizations to Elicit Competing Interpretations},'' \emph{IEEE Transactions on Visualization and Computer Graphics}, vol.~30, no.~1, pp. 1074--1084, 2024.

\bibitem{rosenholtz2011predictions}
R.~Rosenholtz, A.~Dorai, and R.~Freeman, ``{Do Predictions of Visual Perception Aid Design?}'' \emph{ACM Transactions on Applied Perception}, vol.~8, no.~2, pp. 1--20, 2011.

\bibitem{shin2021effects}
D.~Shin, ``The effects of explainability and causability on perception, trust, and acceptance: Implications for explainable ai,'' \emph{International Journal of Human-Computer Studies}, vol. 146, p. 102551, 2021.

\bibitem{reimers2024bars}
S.~Reimers and N.~Harvey, ``Bars, lines and points: The effect of graph format on judgmental forecasting,'' \emph{International Journal of Forecasting}, vol.~40, no.~1, pp. 44--61, 2024.

\bibitem{dehaene2003neural}
S.~Dehaene, ``{The neural basis of the Weber--Fechner law: a logarithmic mental number line},'' \emph{Trends in cognitive sciences}, vol.~7, no.~4, pp. 145--147, 2003.

\bibitem{itti2002model}
L.~Itti, C.~Koch, and E.~Niebur, ``A model of saliency-based visual attention for rapid scene analysis,'' \emph{IEEE Transactions on Visualization and Computer Graphics}, vol.~20, no.~11, pp. 1254--1259, 2002.

\end{thebibliography}


\begin{IEEEbiographynophoto}{Xumeng Wang} is with DISSec in Nankai University. Her research interests are visual analytics and privacy preservation.
\end{IEEEbiographynophoto}

\begin{IEEEbiographynophoto}{Xiangxuan Zhang}is with DISSec in Nankai University. His research focuses on visual analytics and large language models.
\end{IEEEbiographynophoto}

\begin{IEEEbiographynophoto}{Zhiqi Gao} is currently pursuing a bachelor's degree in Information Management and Information Systems at the Business School of Nankai University. He is also serving as a research intern at DISSec at Nankai University. His research interests cover Social Computing and Human-Computer Interaction.
\end{IEEEbiographynophoto}

\begin{IEEEbiographynophoto}{Shuangcheng Jiao} is with DISSec in Nankai University. His work involves visual analytics and privacy preservation.
\end{IEEEbiographynophoto}

\begin{IEEEbiographynophoto}{Yuxin Ma} is a tenure-track Associate Professor in the Department of Computer Science and Engineering, Southern University of Science and Technology (SUSTech), China. His primary research interests are in the areas of data visualization and visual analytics, focusing on applications related to explainable artificial intelligence, high-dimensional data, spatiotemporal data, and interactive education support. His work has been published in various top venues in visualization and human-computer interaction (IEEE TVCG, IEEE VIS, ACM CHI, etc.) and recognized through Honorable Mention Awards at ACM CHI (2022) and CVMJ (2018). He has served as a program committee member and reviewer for major conferences and journals in visualization, human-computer interaction, and artificial intelligence.
\end{IEEEbiographynophoto}

\end{document}